\documentclass[sn-mathphys,Numbered,iicol,referee]{sn-jnl}% Math and Physical Sciences Reference Style
\usepackage{graphicx}%
\usepackage{multirow}%
\usepackage{amsmath,amssymb,amsfonts}%
\usepackage{amsthm}%
\usepackage{mathrsfs}%
\usepackage[title]{appendix}%
\usepackage[dvipsnames]{xcolor}%
\usepackage{textcomp}%
\usepackage{manyfoot}%
\usepackage{booktabs}%
\usepackage{algorithm}%
\usepackage{algorithmicx}%
\usepackage{algpseudocode}%
\usepackage{listings}%

\usepackage{amsmath, amsthm, amsfonts, amssymb, xfrac, mathtools}
\usepackage{bbold}
\numberwithin{equation}{section}
\usepackage{scalerel}  % Makes anything whatever size you want in mathmode
\usepackage{graphicx,color,subcaption}
\usepackage{bm}	
\usepackage[utf8]{inputenc}
\usepackage{soul}
\usepackage[normalem]{ulem}

\usepackage{cancel}
\usepackage[most]{tcolorbox}
\usepackage{empheq} %to put a box around equations
\newtcblisting{myscribbles}{boxrule=0.5pt,fontupper=tiny}
\tcbuselibrary{skins}
\usepackage{tikz,nccmath}
\usetikzlibrary{calc,arrows}

\usepackage{bm}	

\usepackage{cleveref}
\usepackage{syntonly}
\usepackage[multiple]{footmisc}  %Multiple footnotes

\newcommand{\km}{k^-}

\newcommand{\kperp}{\mathbf{k}}

\newcommand{\qperp}{\mathbf{q}}

\newcommand{\pperp}{\mathbf{p}}
\newcommand{\ppperp}{\mathbf{p}'}

\newcommand{\half}{\frac{1}{2}}
\newcommand{\tr}{\text{tr}}
\newcommand{\Tr}{\text{Tr}}
\newcommand{\M}{\mathcal{M}}

\newcommand{\nn}{\nonumber\\}

\newcommand{\etal}{\emph{et al.}\,}

\begin{document}

\title[Article Title]{Gluon radiation from a classical point particle: Recoil effects}


\author[1,2,3]{\fnm{Isobel} \sur{Kolb\'e}}\email{isobel.kolbe@wits.ac.za}

\author[1,4]{\fnm{Mawande} \sur{Lushozi}}\email{mawande.lushozi@uct.ac.za}
\equalcont{These authors contributed equally to this work.}

\affil[1]{\orgdiv{Institute for Nuclear Theory}, \orgname{University of Washington}, \orgaddress{\street{Box 351550,}, \city{Seattle}, \postcode{	98195}, \state{WA}, \country{USA}}}

\affil[2]{\orgdiv{Instituto Galego de Fisica de Altas Enerxias (IGFAE)}, \orgname{Universidade de Santiago de Compostela}, \orgaddress{\street{Rúa de Xoaquín Díaz de Rábago}, \city{Santiago de Compostela}, \postcode{15705 }, \state{Galicia}, \country{Spain}}}

\affil[3]{\orgdiv{School of Physics}, \orgname{University of the Witwatersrand}, \orgaddress{\street{1 Jan Smuts Avenue}, \city{Johannesburg}, \postcode{2000}, \country{South Africa}}}

\affil[4]{\orgdiv{Department of Physics}, \orgname{University of Cape Town}, \orgaddress{\street{University Avenue}, \city{Cape Town}, \postcode{7701}, \country{South Africa}}}


\abstract{The gluon radiation spectrum of a classical particle struck by a sheet of colored glass, is a key ingredient in understanding the  distribution of energy and baryon density in the fragmentation region, particularly in the initial stages of heavy ion collisions. However, the currently known classical spectrum has a troublesome high-momentum tail \cite{Kajantie:2019hft,Kajantie:2019nse}.  
By comparing tree-level bremsstrahlung of a spin-less quark to the above-mentioned known result, we propose an interpolating formula that takes into account the recoil of the struck particle, and therefore produces the correct perturbative behaviour at high momentum of the radiated gluon. 
}

\keywords{Heavy-Ion Collisions, Fragmentation Region, Gluon Bremsstrahlung, pQCD}

\maketitle

\section{Introduction}

    In the context of developing a description of the early stages of heavy-ion collisions, great success has been achieved by describing the initial condition of the collision in terms of the Color Glass Condensate (CGC), followed by a highly coherent phase of matter (called the Glasma), which produces rapid thermalization to the Quark Gluon Plasma \cite{Lappi:2006fp,Gelis:2009wh}.
    The CGC (and subsequent Glasma) picture relies on the realization that the density of gluons in a boosted nucleus becomes so large that their transverse separation becomes negligible.
    Traditionally, this high density of charges is then used to motivate a separation of degrees of freedom that allows one to consider a classical evolution of the small-$x$ modes in the presence of static  large-$x$ color charges.
    Here, the density of color charge is very high, so for sufficiently low transverse resolution, one observes the charge as classical, i.e. belonging to a high-dimensional representation of the color group $SU(3)$.

    The picture described above can be carried over to understand the fragmentation region, that part of phase-space in which the fragments from a collision have rapidity similar to that of the initial projectile or target \cite{Anishetty:1980zp,Kajantie:1982jt,Kajantie:1982nh}.
    In a heavy-ion collider experiment, the fragmentation region corresponds to the very far forward (or backward) region and may offer access to higher density regions of the quantum chromo-dynamical (QCD) phase diagram\cite{Kolbe:2020hem,McLerran:2018avb}.

    A few years ago, McLerran set forth an interesting program to calculate properties of matter produced in the fragmentation region in a manner that generalizes the notion of the Glasma to include finite (net) baryon density \cite{McLerran:2018axu}. Important first steps were taken by Mclerran \etal \cite{McLerran:2018avb}  in calculating the space-time evolution of baryon density, then by Kajantie \etal \cite{Kajantie:2019hft,Kajantie:2019nse} in calculating the resulting gluon radiation classically. 
 The latter result (hereinafter ``the classical result'') is rather remarkable and elegant. However, the authors highlight that the result is only valid in the kinematic region where the recoil of the struck quark may be neglected. Indeed, it does not reproduce the perturbative $\sim 1/k_T^4$ behavior at large gluon transverse momenta $k_T$ (cf. \cite{Dumitru:2001ux}).  In this paper, it is our intention to offer a remedy to the aforementioned shortcoming, in the form of an interpolating formula which is readily useful for phenomenology.

    %In order to advance our understanding of initial energy deposited in the  fragmentation region, particularly in high-energy heavy-ion collisions, we require a realistic estimation of the momentum spectrum of the gluon radiation in this region.  
    %The calculation of the gluon spectrum in the fragmentation region is hampered by the multi-scale nature of the problem, since one must develop an understanding of both the non-perturbative physics of a highly boosted nucleus, as well as the perturbative production of particles.
    
    The literature on gluon radiation, both classically and in perturbative QCD, is extensive, but traditionally deals with computations in the lab frame (see for example \cite{Kovchegov:1998bi,Kopeliovich:1998nw,Kopeliovich:1999am}, at NLO \cite{Li:2021zmf,Li:2021yiv,Li:2021ntt}, and references in \cite{Kajantie:2019hft,Kajantie:2019nse}).
    Extending such calculations to the target's fragmentation region by boosting the kinematics is not straightforward. We do not attempt that here, but instead remain in the target's rest frame as done by Kajantie \etal. When studying the space-time evolution of energy and the matter formed in the target fragmentation region, it can be very helpful to work in the rest frame of the target; it is in this frame that one can clearly understand baryon compression, as well as baryon stopping power in terms of the saturation scale $Q_{\text{sat}}$ \cite{Anishetty:1980zp,Kajantie:1982jt,Kajantie:1982nh,McLerran:2017ep,McLerran:2018avb,McLerran:2018axu,Kajantie:2019hft,Kolbe:2020hem}.

    % but he distribution of gluon radiation in the fragmentation region has been computed in the fully classical limit (hereinafter ``the classical result'') in two papers by Kajantie \etal\cite{Kajantie:2019hft,Kajantie:2019nse}.
   % Those authors highlight that their result is only valid in the kinematic region where the recoil of the struck quark may be neglected.  That is, the fully classical result derived by Kajantie \emph{et al.}\,is only valid in the low-frequency region.	

    %It turns out that the high-frequency limit of the pQCD result is, in fact, different from that of the classical result.  Since the pQCD result is valid in the high-frequency limit and the classical result is not, we are forced to accept that the classical result cannot be correct in the high-frequency limit. 
   % On the other hand, the pQCD calculation is not able to describe the non-perturbative physics of multiple gluon exchanges between the struck quark and the sheet (the boosted nucleus), and so is an incomplete description of the  underlying physics.
   The way in which we arrive at our result builds on the work done in \cite{Lushozi:2019duv}, wherein it is observed that it is also possible to compute an ``analogue" to the classical result in perturbative QCD (pQCD) (hereinafter ``the pQCD result'') in order to account for the recoil of the struck quark. This was done in the hope that a comparison with the classical result would allow one to write down an ansatz for the spectrum of gluon bremsstrahlung that is able to both describe the non-perturbative physics of interactions encompassed in the framework of the Color Glass Condensate (CGC) \cite{McLerran:1993ni,McLerran:2017ep,Kajantie:1982nh}, and produce the correct perturbative behaviour in the high momentum (of the radiated gluon) limit.
   Unfortunately, the inclusion of the physics of spin in the pQCD result \cite{Lushozi:2019duv}, in addition the the fact that the pQCD result is presented at the level of the amplitude squared, obscures any such attempt at comparison.
        
    We solve the problem of reconciling the discordant known perturbative and non-perturbative (classical) results by making a pQCD-inspired modification of the classical result.  
    In order to see more clearly the relationship between the classical and pQCD results, we make two simplifications to the pQCD calculation of \cite{Lushozi:2019duv}, both of which bring the quark closer in nature to the classical point particle studied by Kajantie \etal\cite{Kajantie:2019hft,Kajantie:2019nse,Wong:1970fu}: 
    First, we remove the quark's spin from the problem. Secondly, we treat the quark's color charge as classical (\emph{i.e.}, commuting). We are thus led to study, pertubatively, gluon bremsstrahlung in \emph{scalar QCD in the limit of classical color charge} (we will make this statement more precise in \cref{sec:Scalar} and \cref{App:ClassicalColorCharge}). 
    The result (hereinafter ``the scalar result'') takes a simple form at the amplitude level, and is therefore easy to compare with the classical result, even at the level of the cross-section.  
    We argue that one may simply adjust the form of the classical result to match that of the scalar result.  
    What one is left with is then an ansatz for gluon bremsstrahlung in the fragmentation region (hereinafter ``the ansatz'') that is able to fully take account of non-perturbative interactions, but also produces the correct high-frequency limit.  
    That is, the low-frequency limit of our ansatz is precisely the classical result, while the high-frequency limit falls off in the expected way as $\sim 1/k_T^4$.  
    This paper formalizes the proffered solution and carefully explores the relevant limits.  We also offer a physics interpretation of the connection between the classical result and the ansatz.
    
    What we achieve in the end is a generalisation of the gluon radiation spectrum calculated by Kajantie \etal,to include the effects of the particle's recoil. 
    
    %produced by a classical point particle struck by a sheetof colored glass \cite{McLerran:1993ni,McLerran:1993ka,Iancu:2003xm,Weigert:2005us,Gelis:2010nm}, 
    %This is part of a broader program to understand the fragmentation region of heavy-ion collisions using a CGC-classical framework \cite{Anishetty:1980zp,Kajantie:1982jt,Kajantie:1982nh,McLerran:2017ep,McLerran:2018avb,Kolbe:2020hem}.

    This paper is organised as follows: 
    We start by reminding the reader of the classical result and its properties in \cref{Sec:Class}.
    We will then present the calculation of bremsstrahlung in the fragmentation region within the context of a scalar field with classical color charge in \cref{sec:Scalar}.
    In \cref{sec:GluonRadAnsatz} we present an ansatz for gluon radiation in the fragmentation region by first reformulating the classical result in \cref{sec:RefClass} and then use the scalar result to inform a modification to the classical result in \cref{sec:Ansatz}.
    In \cref{sec:Ansatz} we also spend some time discussing the ansatz before making concluding remarks in \cref{sec:Conclusion}.

\section{Bremsstrahlung in the fragmentation region}

\subsection{Striking a classical color charge with a sheet of colored glass in the no-recoil approximation}\label{Sec:Class}
    
    The distribution of gluon radiation from a classical point particle struck by a sheet of colored glass (representing the nucleus A, associated with a strong background Yang-Mills field $A^\mu$), schematically given by the following process:
		\begin{align}
			A+q(p)\rightarrow X+q(p')+g(k)\,\,,
		\end{align}
	
    has recently been derived in two papers by Kajantie \etal\cite{Kajantie:2019nse,Kajantie:2019hft}, and is given by\footnote{Other bremsstrahlung calculations (for eg. in both the classical and quantum treatments of electrodynamics \cite[Ch. 6]{Peskin:1995ev},\cite[Ch. 1-3-2]{Itzykson:1980rh}, as well as the result presented here in \cref{eq:scalarMult:2}, have an additional factor of $\half$. There appears to be a typo in the final result given in eq. (53) of \cite{Kajantie:2019nse}, which should have 4 on the right-hand-side instead of 2.  This may be seen by substituting eqs. (A13) and (A19) into the unnumbered equation between eqs. (52) and (53) of \cite{Kajantie:2019nse}.}
		\begin{align}
			\left.\frac{dN}{dyd^2k_T}\right\vert_{Cl.}
                    =&\frac{g^2C_F}{16\pi^3}\,4\int \frac{d^2h}{(2\pi)^2} \tilde{S}(\mathbf{k}-\mathbf{h}) \times\nonumber\\
                    &\times\left[\mathcal{M}^i_{\text{Cl.,bulk}}+\mathcal{M}^i_{\text{Cl.,brems}}\right]^2\label{gluon-mult:KMP} 
		\end{align}
    where
		\begin{align}
		    \mathcal{M}^i_{\text{Cl.,bulk}}&= \frac{h^i}{h_T^2+2(k^-)^2}-\frac{k^i}{k_T^2+2(k^-)^2}\,, \\
			\mathcal{M}^i_{\text{Cl.,brems}}&=\frac{k^i}{k_T^2+2(k^-)^2}-\frac{k^i-\xi p^{\prime i}}{|\mathbf{k}-\xi\mathbf{p}'|^2+\xi^2m^2} \,. 
		\end{align}
		
    In \cref{gluon-mult:KMP}, $ \tilde{S}(\mathbf{k}-\mathbf{h})$ is the Fourier Transform of the two-point correlator of two Wilson lines and carries the physics of the scattering of the quark and the projectile nucleus\footnote{Kajantie \etal employ the approximation of the McLerran-Venugopalan (MV) model which will lead to a Gaussian expression for $S(\textbf{x}-\textbf{y})$ \cite{McLerran:1993ni,McLerran:1993ka}.}.  
    The radiated momentum relative to the final momentum of the struck quark is described by the fraction $\xi=\sfrac{k^-}{p'^-}$.  The momentum $\textbf{h}$ arises as a subtlety of convoluting two Fourier transforms and will not play an important role in the present work.

    The calculation of pQCD bremsstrahlung in the fragmentation region has been performed \cite{Lushozi:2019duv} and shows that, as is the case in QED, the gluon multiplicity distribution should fall off as $1/k_T^4$ for large gluon transverse momentum $k_T$. 
    The classical result quoted in \cref{gluon-mult:KMP},  calculated by Kajantie \emph{et al.} \cite{Kajantie:2019nse}, does not display this expected $1/k_T^4$ behavior at high $k_T$, but instead follows a $1/k_T^2$ fall-off throughout. Of the two contributions to their result  $\mathcal{M}_{\text{Cl.,bulk}}$ and $\mathcal{M}_{\text{Cl.,brems}}$, the culprit for the undesired behavior is the bremsstrahlung term $\mathcal{M}_{\text{Cl.,brems}}$. 
    Kajantie \etal  argue that this is because their formulation cannot take quark recoil into account. 
    
    One might argue that there is no reason to believe that the complicated non-perturbative interaction considered by Kajantie \etal should preserve the $1/k_T^4$ behaviour;  However, it must be true that, in the limit that the background field is weak, one must recover the known perturbative result.
    
    In order to incorporate the physics of the recoil, the correct problem to study is bremsstrahlung in ``classical'' scalar QCD\footnote{By classical scalar QCD we mean QCD with spinless quarks and classical (commuting) color charge, see \cref{sec:Scalar}.}.  The spectrum for gluon bremsstrahlung in classical scalar QCD will allow us to quantify the recoil that the fully classical result is unable to.  We will then be in a position to suggest a formula for the spectrum of gluon radiation in the fragmentation region that holds more generally, and reduces to the classical result in the no-recoil approximation.

\subsection{Fully perturbative bremsstrahlung by spin-less classical quarks}\label{sec:Scalar}
	
	We consider a quantum field theory which contains a complex scalar field and a gauge field, both with classical color charge.  We develop the notion of a classical color charge formally in \cref{App:ClassicalColorCharge}, but in practice the term ``classical color charge'' means that the generators $\{T^a\}_{ij}$ of the representation commute.  Such a theory is then a classical and scalar version of QCD, and we will henceforth refer to it as the ``scalar theory''. \textcolor{blue}{}
	
\begin{figure*}
			\centering
			\includegraphics[scale=1]{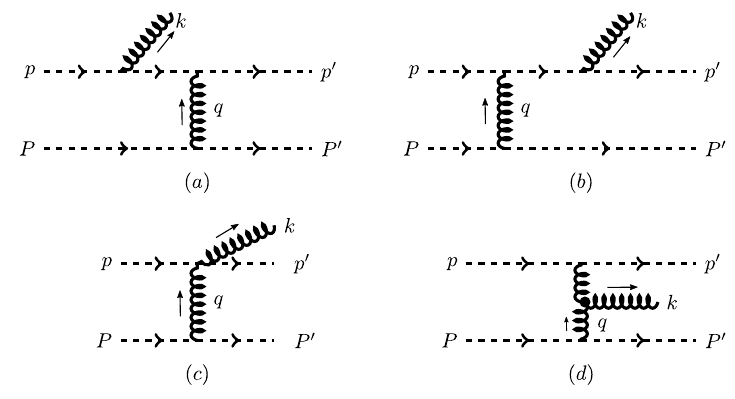}
			\caption{Tree-level diagrams that contribute to bremsstrahlung in a scalar field theory.\label{fig:scalarDiagrams}}
		\end{figure*}

	The relevant Feynman rules of scalar QCD are described in \cref{app:ScFeynRules} and given there in \cref{fig:scalarFeynRules}.  The relevant diagrams for bremsstrahlung at tree level are shown in \cref{fig:scalarDiagrams}. 
	We will compute the diagrams in \cref{fig:scalarDiagrams} using the kinematics that are relevant in the fragmentation region \cite{Lushozi:2019duv}: We will compute the bremsstrahlung resulting from the collision between a `quark' (scalar field) with very high momentum $ P  $ colliding with a stationary `quark' with four-momentum $ p $. 
	We will take the momentum of the lower line $ P $ to be the largest scale in the problem. Starting with the diagrams in \cref{fig:scalarDiagrams}, some algebra will show that the correct kinematics are (in light-cone coordinates $ x^{\mu}=\left[x^+,x^-,\mathbf{x}\right] $)
		\begin{align}
			p^{\mu}	    &=\left[\frac{m}{\sqrt{2}},\frac{m}{\sqrt{2}},\mathbf{0}\right],	\nn
                p'^{\mu}    &=\left[\frac{\xi m_{\perp}^2}{2\km},\frac{1}{\xi}\km,\qperp-\kperp\right],\nn
			P^{\mu}	    &=\left[P,0^-,\mathbf{0}\right],\nn
                P'^{\mu}    &=\left[P-q^+,0^-,-\qperp\right],\nn
			k^{\mu}	    &=\left[\frac{\kperp^2}{2\km},\frac{m}{\sqrt{2}}\,\frac{\xi}{1+\xi},\kperp\right],	\nn
                q^{\mu}     &=\left[\frac{p'^2_{\perp}+m^2}{2m}(1+\xi)+\frac{\kperp^2}{2\km}-\frac{m}{\sqrt{2}},0^-,\qperp\right]. \label{kinematics}
		\end{align}
		
	As usual, in light-cone gauge the polarization vectors must obey $ \epsilon(k)\cdot k=0 $	 and $ n\cdot\epsilon(k)=0 $.  Thus, the two polarization vectors $\epsilon_{\lambda=1,2}^{\mu}$ may be chosen using the standard transverse basis $\{ \hat{e}_1 =(1,0),\hat{e}_2 =(0,1)\}$ so that:
		\begin{equation}
			\epsilon^{\mu}_{\lambda=i}(k)	=\left[\frac{k^i}{\km},0^-,\hat{e}_i\right]\qquad ,\lambda=1,2.\label{polvec}
		\end{equation}
		
	One may check that \cref{polvec} satisfies the conditions $ \epsilon(k)\cdot k=0 $	 and $ n\cdot\epsilon(k)=0 $ using the light-cone metric which defines the dot product $a\cdot b=a^+b^-+a^-b^+-\mathbf{a}\cdot\mathbf{b}$.
		
    The three-gluon diagram (diagram (d) in \cref{fig:scalarDiagrams}) is suppressed because it is proportional to a commutator of the generators. Therefore, the scalar result, at the level of the amplitude, is given by
 
        \begin{align}
    		\mathcal{M}_{\text{Sc.}}
                    =&\scaleto{\epsilon}{6.8pt}^*\!\!\!_{_{ \mu\scaleto{(\lambda)}{5.6pt}}}\!\scaleto{(k)}{10.6pt} \mathcal{M}_{\text{Sc.}}^\mu	\,,\label{eq:Msc}\\
    		\mathcal{M}_{\text{Sc.}}^\mu
                    =&-\frac{g^3}{q_\bot^2}P(T^b)_{mn}\left\lbrace T^b,T^a \right\rbrace_{ij}\times\nn
                    &\times\left[ \frac{(2p-k)^\mu}{2p\cdot k}(p'+p-k)^-\right.\nn
                    &\left.\qquad-\frac{(2p'+k)^\mu}{2p'\cdot k}(p+p'+k)^- +2g^{\mu-}\right]\,\,.\label{eq:MsMu}
		\end{align}
	
	This amplitude is the main ingredient in getting to a multiplicity distribution that takes into account the recoil of the quark. 
	
	It is interesting to compare the expression in the square brackets in the above \cref{eq:MsMu} to the well-known\footnote{See for instance \cite{Jackson:1998nia}, and, in the present context, eq.(86) of \cite{Kajantie:2019hft}.} current associated with classical bremsstrahlung of a charged particle kicked from initial momentum $p$ to $p'$:  
	
    	\begin{align}
    	    J^{\mu}(k)&\sim\frac{p^\mu}{p\cdot k}-\frac{p^{\prime\mu}}{p'\cdot k} \,. \label{classicalJ:brems}
    	\end{align}

	In both \cref{eq:MsMu} and \cref{classicalJ:brems}, the first and second terms correspond to the diagrams in which the bremsstrahlung gluon is radiated before and after the collision respectively.  
	In the first two terms of the fully perturbative, scalar result of \cref{eq:MsMu}, the numerators have two components: (1) the average of the momentum of the upper quark before and directly after radiating the gluon, and (2) the sum of the ``$-$''-momentum of the upper quark before and directly after the collision.
	The classical result in \cref{classicalJ:brems} has the exact same two contributions to the numerators of each term, but is derived in the kinematics that the incoming and outgoing upper quark have the same momentum.  That is, the classical result assumes (i) no recoil, and (ii) low frequency emission, or small $k$.  Assumptions (i) and (ii) imply that (1) is given simply by the incoming or outgoing upper quark momentum respectively, and (2) is factored out in \cref{classicalJ:brems}.  
	Analogously to \cite{Jackson:1998nia}, assumption (ii) of the classical result then also explains why the third term of the scalar result does not have a counterpart in the classical result:  In the limit that $k$ is small, the first two terms of \cref{eq:MsMu} have small denominators, making them large relative to the third term.
	
	We would now like to compute the gluon multiplicity distribution.  A careful derivation of the expression we will use below in \cref{eq:GlMultDist} appears in \cref{app:GlMultDerivation}.  The derivation follows the procedure introduced by \cite{Gunion:1981qs} in which the Born cross section (for the same process without radiation) is used to normalize the cross section before careful evaluation of the phase-space integral.  The resulting expression is as follows.
	    \begin{align}
    	    \left.\frac{dN}{dyd^2k_T}\right\vert_{\text{Sc.}} &= \frac{1}{2}\frac{1}{(2\pi)^3}(1+\xi) \frac{ \vert \mathcal{M}_{\text{Sc.}}\vert^2}{\vert \mathcal{M}_{\text{Sc., Born}}\vert^2}\label{eq:GlMultDist}
    	\end{align}
	
	The appearance of the factor of $(1+\xi)$ is crucial.
	
	We will now use \cref{eq:MsMu} to calculate the gluon multiplicity distribution in \cref{eq:GlMultDist}. 
	To do so we also need to calculate the Born amplitude, given by  
	   	\begin{align}
        	 \mathcal{M}_{\text{Sc., Born}}
        	    &= \frac{g^2}{q_\bot^2}2\sqrt{2}mP(T^a)_{ij}(T^a)_{mn} \,\,.
        \end{align}
        
    Summing over final states and averaging over the initial, we get the following results for the amplitude-squared
        \begin{align}
            \vert&\mathcal{M}_{\text{Sc., Born}} \vert^2
                =\frac{g^4}{q_\bot^4}8m^2P^2\frac{C(R)^2(N_c^2-1)}{d_R^2}\label{eq:BornAmpSq} \,\,,\\
            \vert &\mathcal{M}_{\text{Sc.}} \vert^2
                =   \frac{g^6}{q_\bot^4}P^2\sum\limits_{\lambda}\frac{4C(R)^2C_R(N_c^2-1)}{d_R^2}\times\nn
                &\times\left\vert \frac{4\,\epsilon_\lambda \cdot p}{2p\cdot k}p^{\prime-} 
        			-\frac{4\,\epsilon_\lambda \cdot p'}{2p'\cdot k}(p^{\prime-}+k^-)\right\vert^2  \,\,.\label{eq:Msc-sq}
        \end{align}
	
	Comparing \cref{eq:Msc,eq:MsMu} to \cref{eq:Msc-sq} above, it appears as though we have omitted the term proportional to $\epsilon_\mu g^{\mu-}$. Indeed, this is simply equal to ${\epsilon^{-}}$, which vanishes by our choice of polarization vectors in \cref{polvec}. We have also made use of the kinematics of the problem (\cref{kinematics}) to make the replacement ${p^-\rightarrow (p^{\prime-}+k^-)}$. Here $R$ denotes a representation with dimension $d_R$, chosen to be large enough that the generators essentially commute. We denote the normalization of the inner product by $C(R)$, that is, $Tr(T^aT^b)=C(R)\delta^{ab}$; $C_R$ denotes the quadratic Casimir defined by $T^aT^a=C_R\mathbb{1}$. 
	
	The gluon multiplicity distribution is given by \cref{eq:GlMultDist}:
    	\begin{align}
    	    \left.\frac{dN}{dyd^2k_T}\right\vert_{\text{Sc.}} 
        	    &=\frac{1}{2}\frac{1}{(2\pi)^3}(1+\xi)\frac{g^2C_R}{2m^2} \sum\limits_{i}\times\nn
                    &\kern-5em\times\left\vert \frac{4\,\epsilon_i \cdot p}{2p\cdot k}p^{\prime-} 
        			-\frac{4\,\epsilon_i \cdot p'}{2p'\cdot k}(p^{\prime-}+k^-)\right\vert^2 \,\,.\label{eq:scalarMult:1}
    	\end{align}

	 We  can use the following identities   
    	\begin{align}
        	\scaleto{\epsilon}{6.pt}_{ \scaleto{l}{5pt}}\cdot p                     
                    &=\frac{1+\xi}{\xi} k^l\,\,,\\
                \scaleto{\epsilon}{6.pt}_{ \scaleto{l}{5pt}}\cdot p'  
                    &=\frac{1}{\xi}\left(k^l-\xi p^{\prime l}\right)\,\,,\\
        	2p\cdot k   
                    &= \frac{1+\xi}{\xi}\left(k_T^2+2(k^-)^2\right)\,\, ,\\ 
                2p'\cdot k               
                    &=\frac{1}{\xi} \left(\vert\kperp-\xi\ppperp  \vert^2+\xi^2m^2\right)\,,\\
        	p^{\prime-} 
         &=\frac{1}{1+\xi}p^-=\frac{1}{1+\xi}\frac{m}{\sqrt{2}}\,\,\,\,,
    	\end{align}
	
	derived from \cref{kinematics,polvec}, to rewrite \cref{eq:scalarMult:1} as 
    	\begin{align}
        	\left.\frac{dN}{dyd^2k_T}\right\vert_{\text{Sc.}}	
        	    &=\frac{g^2C_R}{16\pi^3}\,4\, \frac{1}{1+\xi}\times\nn
                    &\kern-5em\times
        	        \left\vert \frac{k^i}{k_T^2+2(k^-)^2}-(1+\xi)\frac{k^i-\xi p^{\prime i}}{\vert\kperp-\xi\ppperp  \vert^2+\xi^2m^2} \right\vert^2\,. \label{eq:scalarMult:2}
    	\end{align}
	
	\Cref{eq:scalarMult:2} is the main result of this paper. It is the lowest order perturbative contribution to gluon bremsstrahlung and includes quark recoil.  The ratio $\xi=k^-/p^{\prime-}$ that features above is a good measure of quark recoil, and we will show how one recovers the classical results of \cite{Kajantie:2019hft} in the no-recoil approximation $\xi\ll 1$ in \cref{sec:Ansatz}.

\section{Ansatz for gluon radiation from a classical particle struck by a sheet of colored glass}\label{sec:GluonRadAnsatz}

\subsection{Reformulating the classical result}\label{sec:RefClass}
    
    We wish to complete the goal of modifying the classical result \cref{gluon-mult:KMP} to a formula that incorporates the effects of quark recoil. We will do so by insisting that the the expansion of \cref{gluon-mult:KMP} to lowest order in the background field should match the functional form of the scalar (and perturbative) result in \cref{eq:scalarMult:2}. 
    
    The expression for the gluon distribution in \cref{gluon-mult:KMP} contains the background field to all orders. All of the background field dependence is carried by $\tilde{S}(\mathbf{k}-\mathbf{h})$ which is the  Fourier transform of the Wilson line correlator $S(\mathbf{x}-\mathbf{y})=\frac{1}{N_c^2-1}\left\langle \Tr\left(\mathcal{U}(\mathbf{x})\mathcal{U}^{\dagger}(\mathbf{y})\right) \right\rangle\!_\rho$  \cite{Kajantie:2019hft}:
		\begin{align}
			\tilde{S}(q_T):=\int d^2re^{-i\mathbf{q}\cdot \mathbf{r}}S(r_T)\,\,,
		\end{align}
with the following normalization
\begin{align}
   \int\frac{d^2q}{(2\pi)^2} \tilde{S}(q_T)&=1\,\,.
\end{align}
	It turns out that $S(r_T)$, in the McLerran-Venugopalan approximation, can be written as an exponential
		\begin{align}
			S(r_T)&=\exp\left(-Q_s^2D(|\mathbf{r}|)\right)\,\,\label{eq:S_Exp}
		\end{align}

	where $D(|\mathbf{r}|)$ depends on the Green's functions of the two-dimensional Laplacian, and $Q_s$ is the saturation momentum, a scale which arises naturally in the CGC framework. 
	Recall that problems involving classical fields correspond to tree level diagrams in perturbation theory. In order to make this connection, we will take the lowest order in the background field of \cref{gluon-mult:KMP} and compare it to lowest order tree-level bremsstrahlung in scalar QCD. 
	
	Expanding \cref{eq:S_Exp} to lowest order we get $\tilde{S}(q_T)\rightarrow (2\pi)^2\delta^{(2)}(\mathbf{q})$. This corresponds to the lowest order in the background field expansion. The gluon multiplicity distribution at lowest order in the background field is therefore given by
		\begin{align}
			\left.\frac{dN}{dyd^2k_T}\right\vert_{\text{Cl.}}&\rightarrow \frac{g^2C_F}{4\pi^3}\times\nn
			    &\kern-5em\times\left[\frac{k^i}{k_T^2+2(k^-)^2}-\frac{k^i-\xi p^{\prime i}}{|\mathbf{k}-\xi\mathbf{p}'|^2+\xi^2m^2}\right]^2\label{eq:bremm_KMP}\,\,.
		\end{align}

\subsection{An ansatz motivated by the scalar result}\label{sec:Ansatz}

	 Comparing \cref{eq:bremm_KMP} to the perturbative result in \cref{eq:scalarMult:2} leads us to propose the following ansatz for gluon bremsstrahlung in the fragmentation region. 
		\begin{align}
			\left.\frac{dN}{dyd^2k_T}\right\vert_{\text{An.}}
			    &=\frac{g^2C_F}{16\,\pi^3}\frac{1}{1+\xi} \,4\int \frac{d^2h}{(2\pi)^2} \tilde{S}(\mathbf{k}-\mathbf{h})\times\nn
			    &\kern-4em\times\left[\frac{h^i}{h_T^2+2(k^-)^2}-(1+\xi)\frac{k^i-\xi p^{\prime i}}{\vert\kperp-\xi\ppperp  \vert^2+\xi^2m^2}\right]^2,\label{eq:KMP-general} \\
			    &\kern-5em=\frac{g^2C_F}{16\pi^3}\,4\int \frac{d^2h}{(2\pi)^2} \tilde{S}(\mathbf{k}-\mathbf{h})\times\nn
			    &\kern-4em\times\left[\mathcal{\widetilde{M}}^i_{\text{An., bulk}}+\mathcal{\widetilde{M}}^i_{\text{An., brems}}\right]^2\,,\label{eq:KMP-general:2}
		\end{align}

	where, in \cref{eq:KMP-general:2}, we have performed a split in the same spirit as is done by Kajantie \etal so that
		\begin{align}
		    \mathcal{\widetilde{M}}^i_{\text{An., bulk}}
                    =&\frac{1}{\sqrt{1+\xi}} \left(\frac{h^i}{h_T^2+2(k^-)^2}\right.\nn
                    &-\left.\frac{k^i}{k_T^2+2(k^-)^2}\right) \\
			\mathcal{\widetilde{M}}^i_{\text{An., brems}}
                    =&\frac{1}{\sqrt{1+\xi}}\left(\frac{k^i}{k_T^2+2(k^-)^2}\right.\nn
                    &-\left.(1+\xi)\frac{k^i-\xi p^{\prime i}}{\vert\kperp-\xi\ppperp\vert^2+\xi^2m^2} \right)\,. \label{eq:AnBremss}
		\end{align}
		
	From the perspective of the diagrams in \cref{fig:scalarDiagrams}, the no-recoil approximation may be expressed as the case in which $p-k\rightarrow p'$, or that $q^{\mu}\rightarrow 0$, which is equivalent to $\xi\ll 1$.  Notice then that, in the no-recoil limit in which $\xi\ll 1$, the ansatz in \cref{eq:KMP-general} reduces to classical result in \cref{gluon-mult:KMP}.

    To see that \cref{eq:KMP-general:2} indeed cures the undesired high-$k_T$ behavior of the classical result,
    we will separate the bremsstrahlung, bulk, and interference contributions to the gluon radiation distribution, respectively given by
     \onecolumn
        \begin{align}
           	\left.\frac{dN}{dyd^2k_T}\right\vert_{\text{An., brems}}&=\frac{g^2C_F}{4\pi^3}\int \frac{d^2h}{(2\pi)^2} \tilde{S}(\mathbf{k}-\mathbf{h}) \left\vert  \mathcal{\widetilde{M}}^i_{\text{An., brems}} \right\vert^2 \label{eq:bremsrad}\\
            \left.\frac{dN}{dyd^2k_T}\right\vert_{\text{An., bulk}}&=\frac{g^2C_F}{4\pi^3}\int \frac{d^2h}{(2\pi)^2} \tilde{S}(\mathbf{k}-\mathbf{h}) \left\vert  \mathcal{\widetilde{M}}^i_{\text{An., bulk}} \right\vert^2 \label{eq:bulkrad}\\
            \left.\frac{dN}{dyd^2k_T}\right\vert_{\text{An., int}}&=\frac{g^2C_F}{2\pi^3}\int \frac{d^2h}{(2\pi)^2} \tilde{S}(\mathbf{k}-\mathbf{h}) \left(\mathcal{\widetilde{M}}^i_{\text{An., bulk}} \right) \left(\mathcal{\widetilde{M}}^i_{\text{An., brems}} \right).\label{eq:interfrad}
        \end{align}
    
    It is helpful to express the above contributions in terms of the variables $\mathbf{q}=\mathbf{k}+\mathbf{p}'$ and $x=k^-/p^-$, giving
    \begin{align}
		    \mathcal{\widetilde{M}}^i_{\text{An., bulk}}&=\sqrt{1-x} \left(\frac{h^i}{h_T^2+m^2x^2}-\frac{k^i}{k_T^2+m^2x^2}\right)\label{eq:MBulk} \\
			\mathcal{\widetilde{M}}^i_{\text{An., brems}}&=\sqrt{1-x}\left(\frac{k^i}{k_T^2+m^2x^2}-\frac{k^i-xq^i}{|\mathbf{k}-x\mathbf{q}|^2+m^2x^2} \right)\\
			&=\sqrt{1-x}\left(\frac{k^i}{k_T^2+m^2x^2}-\frac{k^i}{|\mathbf{k}-x\mathbf{q}|^2+m^2x^2} +\frac{xq^i}{|\mathbf{k}-x\mathbf{q}|^2+m^2x^2}\right)\,.\label{eq:Mbrems} 
    \end{align}
    
    For $k_T\gg x\, q_T$, one can see that  the first two terms in \cref{eq:Mbrems} will cancel and $\widetilde{M}^i_{\text{An., brems}}\sim 1/k_T^2$. Since  $\widetilde{M}^i_{\text{An., brems}}$ in \cref{eq:bremsrad} does not depend on the integration variable, and the integral of $\tilde{S}$ gives unity, it follows that the bremsstrahlung contribution, \cref{eq:bremsrad},  falls off as $\tilde1/k_T^4$. To study the interference term, \cref{eq:interfrad}, we shift the integration variable $h\rightarrow h+k$, giving
    
        \begin{align}
            &\left.\frac{dN}{dyd^2k_T}\right\vert_{\text{An., int}}=\frac{g^2C_F}{2\pi^3}\int d^2h\,\tilde{S}(\mathbf{k}-\mathbf{h})\mathcal{M}^i_{\text{An., bulk}}\mathcal{\widetilde{M}}^i_{\text{An., brems}}\nonumber\\
            &=\frac{g^2C_F}{2\pi^3}\mathcal{\widetilde{M}}^i_{\text{An., brems}}\int d^2h\,\tilde{S}(-\mathbf{h})\left( \frac{h^i +k^i}{|\mathbf{h}+\mathbf{k}|^2+m^2x^2} -\frac{k^i}{k_T^2+m^2x^2} \right)\\
            &=\frac{g^2C_F}{2\pi^3}\mathcal{\widetilde{M}}^i_{\text{An., brems}}\int d^2h\,\tilde{S}(-\mathbf{h})\left( \frac{h^i}{|\mathbf{h}+\mathbf{k}|^2+m^2x^2}+\frac{k^i}{|\mathbf{h}+\mathbf{k}|^2+m^2x^2} -\frac{k^i}{k_T^2+m^2x^2} \right)\,\,. \label{int:bulk} 
        \end{align}
    \twocolumn
    
    Since the integral over $h$ in \cref{int:bulk} has an upper cut-off set by $\sim Q_s$, one observes that, for $k_T\gg Q_s$, the last two terms in the parentheses cancel and the remaining part of the integrand goes as $1/k_T^2$. So for $k_T$ large enough, both $\widetilde{M}^i_{\text{brems}}$, outside the integral, and the integral itself fall off as $1/k_T^2$, therefore the interference term, \cref{eq:interfrad}, falls off as $1/k_T^4$.
    
    Using the same shift of integration variable, it is clear that the bulk contribution, \cref{eq:bulkrad}, also falls of as $1/k_T^4$ for $k_T\gg Q_s$. We therefore conclude that the ansatz gluon multiplicity distribution \cref{eq:KMP-general:2}, being the sum of all three contributions, has the expected $1/k_T^4$ fall-off at large $k_T$.   
    
        \begin{figure*}
           \centering
           \includegraphics[scale=0.8]{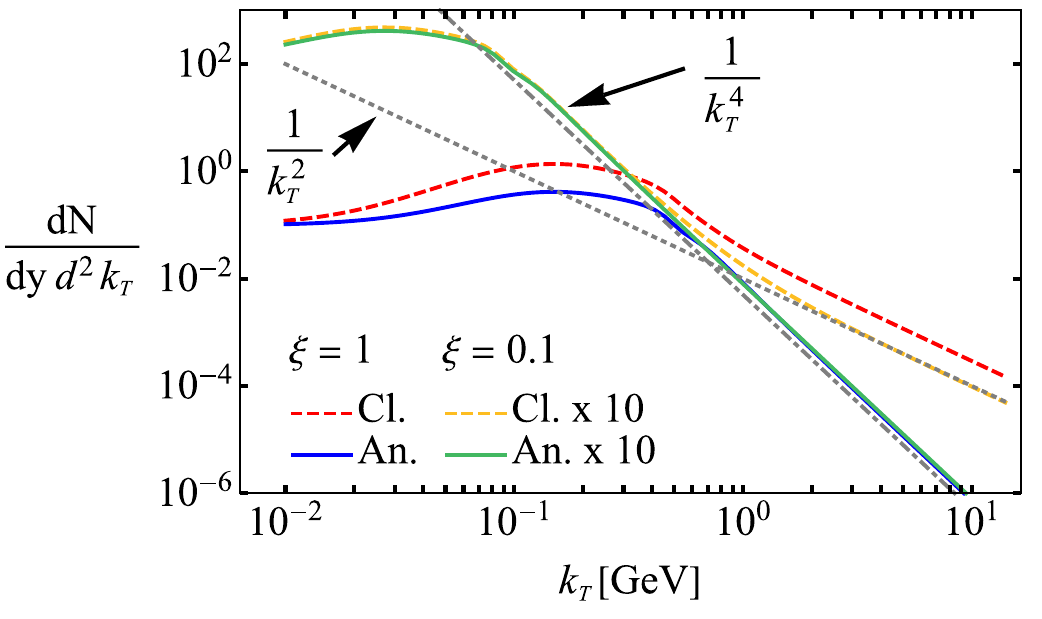}
           \caption{The bremsstrahlung contribution (to lowest order in the background field) to the gluon spectrum in the classical result \cref{eq:bremm_KMP} (dashed red and orange, ``Cl.''), and in the proposed ansatz \cref{eq:AnBremss} (solid blue and green, ``An.'' ), as a function of the radiated gluon momentum $k_T$ for $\xi=1$ (red and blue) and $\xi=0.1$ (orange and green).  Also shown are $\sim\frac{1}{k_T^2}$ (gray dotted) and $\sim\frac{1}{k_T^4}$ (gray dot-dashed) curves to guide the eye. \label{fig:kTBehavior}}
        \end{figure*}

    As an illustration of the large-$k_T$ behavior, we present \cref{fig:kTBehavior}, showing the bremsstrahlung contribution to the gluon spectrum, comparing the classical result with the ansatz.  
    The curves in \cref{fig:kTBehavior} show the gluon multiplicity to lowest order in the background field (as described in \cref{sec:RefClass}) so that the dashed red and orange curves are given by \cref{eq:bremm_KMP} and the solid blue and green curves are given by a similar expression with the square parenthesis as in \cref{eq:KMP-general:2}.
    It is important to note that $\textbf{p'}=\textbf{q}-\textbf{k}$, with $q_T=\sqrt{\qperp\cdot\qperp}$ set by the saturation scale $q_T=Q_s \sim 1$ GeV.  We have further taken the mass of the struck quark to be $m=0.3 $ GeV and plotted for $\xi=1$ and $\xi=0.1$.  We have also integrated the angular dependence.
    \Cref{fig:kTBehavior} shows the two major properties of our ansatz: (1) the ansatz (solid curves) exhibits the correct $k_T^{-4}$ behaviour at large $k_T$ while the classical result (dashed curves) goes like $k_T^{-2}$ for large $k_T$; (2) for values of $\xi\ll1$ the classical result and the ansatz agree in the low-$k_T$ region where the classical result is to be trusted.

\section{Conclusion}\label{sec:Conclusion}
	
    In this work we have presented a formula for the gluon bremsstrahlung spectrum from a single target quark struck by a sheet of colored glass. The extension to the  fragmentation region of a nucleus-nucleus collision can be carried out in a straight forward manner as shown in \cite{Kolbe:2020hem}. We now have the final puzzle piece for the initial conditions of the fragmentation region of heavy-ion collisions, using the classical framework laid out by Kajantie \etal \cite{Kajantie:2019hft,Kajantie:2019nse}.
    
    Our formula utilizes both the correct non-perturbative, low-$k_T$ physics, inspired by the work of Kajantie \etal \cite{Kajantie:2019hft,Kajantie:2019nse}, as well as the correct perturbative high-$k_T$ behaviour, inspired by the result for gluon bremsstrahlung by a classical scalar field, presented here in \cref{sec:Scalar}.

    In order to arrive at an ansatz for the spectrum of gluon radiation in the fragmentation region, we have used the form of the perturbative scalar result, at the amplitude level, to 
    inspire an appropriate modification of the associated classical result. 
    This approach, with a little more work, could possibly allow for the inclusion of spin effects. It would be interesting to see how the radiation formula is modified in that case.
   
\section{Acknowledgements}

    IK and ML were partially supported by the U.S. DOE under Grant No. DE-FG02-00ER41132, and partially under the Multifarious Minds grant provided by the Simons Foundation.   
    The authors also wish to thank Keijo Kajantie, Larry McLerran, Risto Paatelainen, and Michal Praszalowicz for valuable discussions.

\begin{appendices}

\section{Classical Color Charge}\label{App:ClassicalColorCharge}
    
    We would like to formalize what we mean by ``classical color charge''. First note that the usual identities (see for instance \cite{Peskin:1995ev}) may be generalized to an arbitrary representation $ R $ of $ SU(3) $ with dimension $ d_R $.  For instance, there are two casimir operators $ C_R $ and $ C(R) $:
    	\begin{enumerate}
    		\item ``Normalization Casimir'': $ \tr[T^aT^b] = C(R)\delta^{ab} $.
        		\begin{alignat}{2}
        			C(N)	&=\half		&& \qquad\text{(Fundamental rep.)}	\nn
                                &           &&\overset{SU(3)}{\longrightarrow} C(N)=\half	\nn
        			C(G)	&=N			&& \qquad\text{(Adjoint rep.)}		\nn
                                &           &&\overset{SU(3)}{\longrightarrow} C(G)= 3
        		\end{alignat}
    		\item ``Two-Casimir'': $ T^aT^a=C_2(R)\mathbb{1}\equiv C_R\mathbb{1} $	
        		\begin{alignat}{2}
        			C_2(N)	&=\frac{N^2-1}{2N}		&&\qquad \text{(Fundamental rep.)}	\nn
                                &                       &&\overset{SU(3)}{\longrightarrow} C_2(N)=C_F=\frac{4}{3}	\nn
        			C_2(G)	&=N			&&\qquad \text{(Adjoint rep.)}		\nn
                                &                       &&\overset{SU(3)}{\longrightarrow} C_2(G)=C_A= 3
        		\end{alignat}				
    	\end{enumerate}
    	
    We will use, from \cite{Peskin:1995ev} (A.35) (where $ d_G=d_F^2-1=N_c^2-1 $)
    	\begin{align}
    		C(R)	&=\frac{d_R}{d_G}C_R,
    	\end{align}	
    and from \cite{Peskin:1995ev} (A.36)
    	\begin{align}
    		T^aT^bT^a	
    			=&\left[C_2(R)-\half C_2(G)\right]T^b\qquad	\nn
                    &\overset{SU(3)}{\longrightarrow}
    				\left[C_R-\half C_A\right]T^b \label{eq:Peskin_A.36.1}\\
    		f^{abc}T^bT^c
    			=&\half i C_2(G)T^a						\nn
                    &\overset{SU(3)}{\longrightarrow} \half i C_AT^a
    	\end{align}
    
    We consider a charge described by the generator of a representation of $  SU(3) $ that has dimension so large that the generators essentially commute:
    	\begin{align}
    		T^aT^b&=\frac{1}{2}\lbrace T^a,T^b\rbrace+\frac{1}{2}[T^a,T^b]\simeq\frac{1}{2}\lbrace T^a,T^b\rbrace\,\, .
    	\end{align}
    	
    Formally, one may write
    	\begin{align}
    		&\left\Vert \left\lbrace T^a,T^b \right\rbrace\right\Vert
    			\gg \left\Vert \left[T^a,T^b\right] \right\Vert\\
    		\implies& Tr(\lbrace T^a,T^b\rbrace\lbrace T^b,T^a\rbrace  )
    			\gg Tr([T^a,T^b][T^b,T^a])\\
    		2Tr&(T^aT^bT^aT^b)+2Tr(T^aT^aT^bT^b)\nn
    			&\gg i^2f^{abc}f^{bad}Tr(T^cT^d)\label{traces}
    	\end{align}
    
    If we sum over $a$ and $b$ then the traces above  become
    	\begin{align}
    		2Tr(T^aT^aT^bT^b)
    			&=2C_RTr(T^bT^b)\nonumber\\
    			&= 2d_RC_R^2\label{eq:aabb}\\
    	\intertext{and}
    		2Tr(T^aT^bT^aT^b)
    			&=2Tr(T^aT^aT^bT^b)-C(R)f^{abc}f^{abc}\nonumber\\
    			&=2d_RC_R^2-C(R)f^{abc}f^{abc}\,\,.\label{eq:abab}
    	\end{align}
    
    Putting these into \cref{traces} we get
    	\begin{align}
    		2d_RC_R^2
                    &-C(R)f^{abc}f^{abc}+ 2d_RC_R^2
                        \gg C(R) f^{abc}f^{abc}\\
    		      &\implies  2d_RC_R^2\gg C(R) f^{abc}f^{abc}\,\,.\label{color-factor-gg}
    	\end{align}
    
    Also this means
    	\begin{align}
    		Tr(\lbrace T^a,T^b\rbrace\lbrace T^b,T^a\rbrace )
    			&=4d_RC_R^2-C(R)f^{abc}f^{abc}\nn
    			&\simeq 4d_RC_R^2\,,\label{trace-acomm}
    	\end{align}
    Where we  drop $C(R)f^{abc}f^{abc}$ by \cref{color-factor-gg}.

\section{Feynman rules for scalar quarks carrying classical color charge}\label{app:ScFeynRules}

		\begin{figure*}
			\centering
			\includegraphics[scale=0.8]{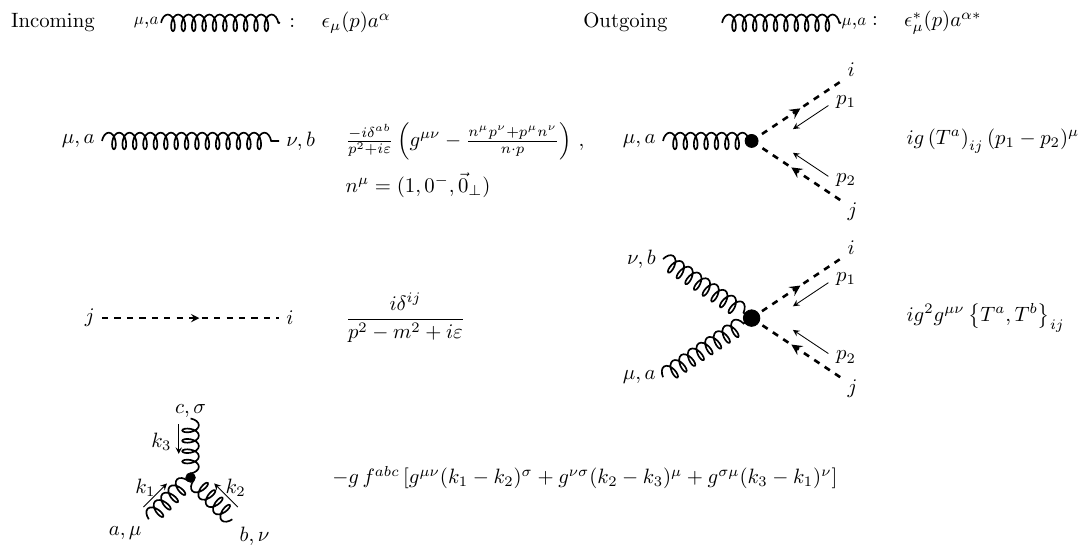}
			\caption{The Feynman rules for a scalar field theory with classical charge.\label{fig:scalarFeynRules}}
		\end{figure*}
	
	In order to derive the Feynman rules for scalar quarks that carry classical color charge, we start with a non-interacting Lagrange density $\mathcal{L}_0$
        \begin{equation}
            \mathcal{L}_0 = \left(\partial^{\mu}\phi\right)^{\dagger}\partial_{\mu}\phi -m^2\phi^{\dagger}\phi,
        \end{equation}
        
    which is invariant under global SU(N) transformations $U=e^{i\alpha T^{a}}$.  We then elevate to a local gauge theory in the usual way by introducing $\partial^{\mu}\rightarrow D^{\mu}=\partial^{\mu}+igT^a A^{\mu a}$, so that
        \begin{align}
            \mathcal{L}_0\rightarrow\mathcal{L}
                =&\left(\partial^{\mu}\phi\right)^{\dagger}\partial_{\mu}\phi
                    +igA_{\mu}^a(T^a)_{i,j}\times\nn
                &\kern-2em\times\left[
                        \left(\partial^{\mu}\phi^{\dagger}_i\right)\phi_j\phi_i^{\dagger}\left(\partial_{\mu}\phi_j\right)\right]\nn
                &+g^2\left(T^aT^b\right)_{ij}A^{\mu a}A_{\mu}^b\phi_i^{\dagger}\phi_j +m^2\phi_i^{\dagger}\phi_i.\label{eq:ScL}
        \end{align}
    
    We will take the generators $T^a$ in \cref{eq:ScL} to be the generators of a representation with a very large dimension, as in \cref{App:ClassicalColorCharge}, describing a classical color charge.
    
    From \cref{eq:ScL} we arrive at the Feynman rules for a scalar field theory carrying classical color charge, given in \cref{fig:scalarFeynRules}.

\section{Derivation of the gluon multiplicity distribution}\label{app:GlMultDerivation}

    In this section we will derive \cref{eq:scalarMult:1}, the expression for the gluon multiplicity distribution.
    We consider the process $ p^{\mu} +P^{\mu} \rightarrow p'^{\mu}+P'^{\mu}+k^{\mu} $ with $ q^{\mu}$ the exchange momentum.
	
	We start by defining the gluon multiplicity distribution $ \frac{dN}{d^2\kperp \,dy} $ in terms of a suitable normalization of the cross section $ \frac{d\sigma}{d^2\qperp\,dy} $ for the process.  We will take the Born cross section (the same scattering without radiation) to define the normalization, as in \cite{Gunion:1981qs}:
		\begin{equation}
			\frac{d\sigma}{d^2\kperp \,dy}
				=\int d^2\qperp\frac{d\sigma_{\text{Born}}}{d^2\qperp} \frac{dN}{d^2\kperp\, dy}.\label{eq:MultDistDef}
		\end{equation}
	
	We need to compute the two cross sections.  In the usual manner, the Born cross section is given by
		\begin{align}
			\frac{d\sigma_{\text{Born}}}{d^2\qperp}
				&=\frac{1}{(2\pi)^2}\frac{1}{flux}\frac{1}{4 P M} \left\vert\mathcal{M}_{\text{Born}}\right\vert^2,\label{eq:XsecBorn}
		\end{align} 
	
	where $ P =P^+$ and $ M=\sqrt{2}\,m $ for $ m $ the quark mass. The $ flux $ factor will cancel.
	
	The bremsstrahlung cross section requires more work as the phase-space integral is non-trivial.  
	The expression that needs to be computed is
		\begin{align}
			d\sigma
				&=\int\frac{1}{(2\pi)^5}\frac{1}{flux}\left\vert\mathcal{M}\right\vert^2
					\frac{d^3P'}{2E_{P'}}\delta_+\left(p'^2-m^2\right)
					\frac{d^3k}{2E_k}.\label{eq:xsec1}
		\end{align}
		
	Now, notice that, since $ P'=P-q $, so that we may perform a change of variables
		\begin{align}
			d^3P'
				&=d^4P'\delta^{(4)}\left(\vec{P'}^2-m_{P'}^2\right)\Theta(P')\\
				&=d^4(P-q)\delta^{(4)}\left((P-q)^2-m_{P'}^2\right)\\
				&=\int dq^+dq^-d^2\qperp\,\delta^+\left(\left(P-q\right)^2\right),\label{eq:d3P'}
		\end{align}
	
	since $ m_{P'}\simeq 0 $ (because $ P'$ is the largest scale), and $ P' $ has only a ``+''-component.  Since the derivative of $ x^2 $ is 0 at $ x=0 $, the delta function in \cref{eq:d3P'} is poorly defined. However, the relevant part of \cref{eq:xsec1} is, in fact,
		\begin{align}
			\int
                    &\frac{d^3P'}{2E_{P'}}\delta^+(p'^2-m^2)\nn
				&=\frac{1}{2E_{P'}}\int dq^+\,dq^-\,d^2\qperp\,\delta^+\left((P-q)^2\right)\,\delta^+(p'^2-m^2).\label{eq:ps1}
		\end{align} 
	
	We may do a change of variables for the second delta function, using the kinematics, given in \cref{kinematics}:
		\begin{align}
			p'^2-m^2
				=&2\left(
					\frac{m}{\sqrt{2}}+q^+-\frac{\kperp^2}{\sqrt{2}m}\frac{1}{x}\right) 
					\frac{m}{\sqrt{2}}(1-x)\nn
                    &-(\qperp-\kperp)^2-m^2\\
				=&\sqrt{2}m(1-x)q^+
					+m^2(1-x)-\kperp^2\frac{1-x}{x}\nn
                    &-(\qperp-\kperp)^2 -m^2.\label{eq:nbqp}
		\end{align}
	
	\Cref{eq:nbqp} is a function of $ q^+ $, and $ p'^2-m^2 =0$ has a root at $ q^+_0 $:
		\begin{equation}
			q^+_0
				=\frac{-m^2(1-x)+\kperp^2(\frac{1-x}{x})+(\qperp-\kperp)^2-m^2}{\sqrt{2}m(1-x)}.
		\end{equation}

	Now, using the identity
	
		\begin{align}
			\delta\left(g(x)\right)
				&=\sum_i\frac{\delta(x-x_i)}{\vert g'(x_i)\vert},
		\end{align}
	
	for $ x_i $ the roots of $ g(x)=0 $, with $ g(x) $ the right hand side of \cref{eq:nbqp}, we may write \cref{eq:ps1} as
		\begin{align}
			\int\frac{d^3P'}{2E_{P'}} \delta^+(p'^2-m^2)
				&=\int\frac{1}{2E_{P'}}\int d^2\qperp\,
					\frac{1}{\sqrt{2}\,m(1-x)}\nn
				&=\int d^2\qperp\,\frac{1}{4PM}\frac{1}{1-x}.\label{eq:d3P'2}
		\end{align}
	
	The last element of \cref{eq:xsec1} is
		\begin{align}
			\frac{d^3k}{2E_k}
				&=\half \frac{dx}{x}d^2\kperp = \half \,dy \,d^2\kperp.\label{eq:d3k}
		\end{align}
	
	Substituting \cref{eq:d3P'2,eq:d3k} into \cref{eq:xsec1} gives
		\begin{align}
			d\sigma	
				&=\frac{1}{(2\pi)^5}\frac{1}{flux}
					\int d^2\qperp\,dy\,d^2\kperp\,
					\frac{1}{4PM}\frac{1}{1-x}\half \left\vert\mathcal{M}\right\vert^2\nn
			\frac{d\sigma}{dy\,d^2\kperp}
				&=\frac{1}{(2\pi)^5}\frac{1}{flux}\frac{1}{8PM}\frac{1}{1-x}
						\int d^2\qperp\left\vert\mathcal{M}\right\vert^2
		\end{align}
		
	We now have all the ingredients to compute the gluon multiplicity distribution \cref{eq:MultDistDef} by substituting in \cref{eq:XsecBorn} and \cref{eq:d3P'2}:
		\begin{align}
			\frac{dN}{d^2\kperp\, dy}
				&=\frac{1}{(2\pi)^3}\half\frac{1}{1-x}
					\frac{\left\vert\mathcal{M}\right\vert^2}{\left\vert\mathcal{M}_{\text{Born}}\right\vert^2}.
		\end{align}
	
	Finally, since $ x=\frac{\xi}{1+\xi} $, we arrive at the expression for the gluon multiplicity distribution used in the current manuscript:
		\begin{align}
			\frac{dN}{d^2\kperp\, dy}
			&=\frac{1}{(2\pi)^3}\half(1+\xi)
			\frac{\left\vert\mathcal{M}\right\vert^2}{\left\vert\mathcal{M}_{\text{Born}}\right\vert^2}.
		\end{align}

\section{Universal Notation}\label{App:UniversalNotation}

\subsection{Bremsstrahlung in perturbative QCD}\label{App:LMPYBremm}
	    
\begin{figure*}
    			\centering
    			\includegraphics[scale=1]{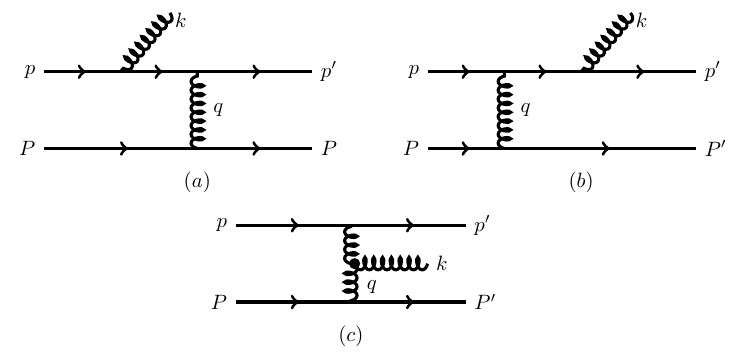}
    			\caption{Tree level diagrams to be computed in the QCD calculation. \label{fig:pQCDTree}}
    		\end{figure*}

	    Lushozi \etal\cite{Lushozi:2019duv} have computed the process $qq\rightarrow qqg$ to lowest order in perturbative QCD in the fragmentation region.  For ease of reference, the diagrams used in the calculation by Lushozi \etal are reproduced in \cref{fig:pQCDTree}. Lushozi \etal consider a quark with very high momentum $ P $ striking a stationary quark and inducing bremsstrahlung, in the kinematic limit of the momentum $ P $ being the largest scale in the problem\footnote{As an aside, one may also show that diagrams with emissions from the bottom, very high-momentum quark are suppressed by a factor of $ P $. }. In order to make the relevant comparison, we will have to rewrite the result in \cite{Lushozi:2019duv} by taking the classical limit as described in \cref{App:ClassicalColorCharge}. 
	
		With a classical color charge, the squares of the remaining two diagrams ((A) and (B)) have the same color factor\footnote{The color factor is not immediately obvious from the form of the result in \cite{Lushozi:2019duv} since the prefactors in those results already include the averaging over initial states, which is to say that the amplitudes in equations (13), (14), and (16) of \cite{Lushozi:2019duv} have already been divided by a factor of $ 2 d_R^2 = 2 N_c^2 $ for quarks.}:  
			\begin{align}
				\left\vert c\left(\M_A\right)\right\vert^2
					&=\left\vert c\left(\M_B\right)\right\vert^2	\\
					&=\left[T^aT^b\right]_{j_1,i_1}T^b_{j_2,i_2}T^{b'}_{i_2,j_2}\left[T^{b'}T^a\right]_{i_1,j_1}	\\
					&=Tr\left(T^aT^bT^{b'}T^a\right)Tr\left[T^bT^{b'}\right]	\\
					&=Tr\left(T^aT^bT^{b'}T^a\right)C(R)\delta^{b b'}	\label{eq:thirdToLast}		\\
					&=C_R^2C(R)\,Tr\left(\mathbb{1}\right)				\label{eq:secondToLast}		\\
					&=C_R^2C(R)d_R.
			\end{align}

        Line \cref{eq:thirdToLast} is by the definition of $ C(R) $, and line \cref{eq:secondToLast} by \cref{eq:aabb}.
		A similar calculation will lead to 
			\begin{align}{2}
				c(\M_A)&\,c^*(\M_B)
					=C_R^2C(R)d_R\left(1-\half\frac{C_A}{C_R}\right)	\\
					&\simeq C_R^2C(R)d_R								\qquad,(C_R\gg C_A) 
			\end{align}
		
		We have used $R$ to denote the representation we have chosen (in which the generators commute), so that $d_R$ is the dimension of the representation, $C_R$ is the usual Casimir (defined by $T^aT^a=C_R\mathbb{1} $, such that $C_F=\frac{4}{3}$, $C_A=3$) and $C(R)$ is the normalization Casimir (defined by $Tr(T^aT^b)=C(R)\delta^{ab}$, such that $C(F)=\half$ and $C(A)=3$).  In the limit of classical color charge then, the squares of the amplitudes of the two contributing diagrams in \cref{fig:pQCDTree}, along with the corresponding ``Born'' diagram (the case of no radiation) are given by:

        \onecolumn
			\begin{align}
				\left\vert\M_A^{QCD}\right\vert^2	
					&=4 K\frac{C_R^2C(R)}{d_R}\left[\frac{x^2}{D_A}+\frac{4x^2M^2(x-1)}{D_A^2}\right]\\
				\left\vert\M_B^{QCD}\right\vert^2	
					&=4 K\frac{C_R^2C(R)}{d_R}\left[\frac{x^2}{D_B}+\frac{4x^2M^2(x-1)}{D_B^2}\right]\\
				\left\vert\M_A^{QCD}\M^{QCD,*}_B\right\vert	
					&=2 K\frac{C_R^2C(R)}{d_R}\left[-\frac{x^2}{D_B}+\frac{x^2(x^2-2c+2)}{D_AD_B}\qperp^2
						-\frac{x^2}{D_A}+\frac{8x^2M^2(1-x)}{D_AD_B}\right]\\
				\left\vert\M_{Born}\right\vert^2
					&=\frac{K}{g_s^2(1-x)}\frac{C(R)C_R}{d_R}
			\end{align}
		
		Where $D_A=\kperp^2 +2x^2M^2$, $D_B = \left(\kperp-x\qperp\right)^2+2x^2 M^2$, $K=8g_s^2(1-x)M^2P^2/q_T^4$, $2 M^2=m^2$, and $m$ is the quark mass.  Therefore the square of the sum of the amplitudes is given by
			\begin{align}
				\left\vert\M^{\text{QCD}}_A\right\vert^2 
					&+\left\vert\M^{\text{QCD}}_B\right\vert^2+ 2\left\vert\M^{\text{QCD}}_A\M_B^{\text{QCD},*}\right\vert\nn\\
					&=\frac{4\,K}{d_R}C_R^2C(R)
						\Bigg[
						4x^2(x-1)M^2\left(\frac{1}{D_A}-\frac{1}{D_B}\right)^2+\frac{x^2(x^2-2x+2)}{D_AD_B}q_T^2\,,
						\Bigg]
			\end{align}
		
		Following the methods of \cite{Lushozi:2019duv} closely again, we find that the gluon spectrum for bremsstrahlung in perturbative QCD, in the limit of classical color charge, is given by
			\begin{align}
				\left.\frac{dN}{d^2\kperp\, dy}\right\vert_{QCD}
					&=\half\frac{1}{(2\pi)^3} g_s^2 C_R x^2\left[4(x-1)M^2\left(\frac{1}{D_A}-\frac{1}{D_B}\right)^2
						+\frac{x^2-2x+2}{D_AD_B}q_T^2\right].
			\end{align}
		
		Note that this is precisely the part of the full result given in eq. (21) of \cite{Lushozi:2019duv} that is proportional to $ C_F $.	
\twocolumn		
\subsection{Fully classical result}\label{App:RewriteKMP}
	
		One may also go through the rather tedious task of writing the classical result \cref{eq:bremm_KMP} in terms of the variables of the perturbative result.  Note that the directions are reversed between the perturbative and classical results (so that the minus direction in one is the plus direction in the other).  Since $x\equiv\frac{k^+}{p'^+} $ in the perturbative calculation, we make the association that
    		\begin{align}
    			\xi=\frac{x}{1-x}.
    		\end{align}
    	
    	We may then, realizing that we must also make the replacement $k^-\leftrightarrow k^+$, further make associations such as
    	    \begin{align}
    	        D_A &= \kperp^2+2(k^+)^2\,,\\
    	        D_B &= \left(\frac{1}{1+\xi}\right)^2\big[(\kperp-\xi \pperp)^2+\xi^2m^2\big]\,.
    	    \end{align}
    	
    	and therefore that the square of the bremsstrahlung term in the classical result may be written as
    		\begin{align}
    			\left\vert\M^i_{\text{bremm.}}\right\vert^2_{Cl.}  
    			     =&-2x^2M^2\left(\frac{1}{D_A}+\frac{1-x}{D_B}\right)^2\nn
                        &\kern-3em+\frac{x}{D_A}-\frac{x(1-x)}{D_B}
    			+x^2(1-x)\frac{q_T^2}{D_AD_B}.\label{eq:ClassAmpSq}
    		\end{align}

\end{appendices}


\bibliography{sn-bibliography}% common bib file

\end{document}